\documentclass[11pt,a4paper]{article}
\usepackage[utf8]{inputenc}
\usepackage{amsmath,amsfonts,amssymb,amsthm,textcomp}
\usepackage{amsthm}
\usepackage[font=footnotesize,labelsep=period]{caption}
\usepackage{float}
\usepackage{multirow}
\usepackage{multicol}
\usepackage{graphicx}
\usepackage[margin=2.5 cm]{geometry}
\usepackage{titlesec}
\titlelabel{\thetitle.\quad}
\usepackage[all]{xy}
\usepackage{hyperref}
\hypersetup{
    colorlinks=true,
    linkcolor=red,
    citecolor=blue
}

\newtheorem{theorem}{Theorem}[section]

\newtheorem{corollary}{Corollary}[theorem]

\begin{document}

\centerline{\Large \bf Miura-reciprocal transformation and symmetries}
\vskip 0.35cm
\centerline{\Large \bf for the spectral problems of KdV and mKdV}
\vskip 0.7cm

\centerline{P. Albares and P. G. Estévez}
\vskip 0.5cm
\centerline{Departamento de F\'isica Fundamental, Universidad de Salamanca,}
\centerline{37008 Salamanca, Spain}

\vskip 0.7cm

\begin{abstract}
\noindent
We present reciprocal transformations for the spectral problems of Korteveg de Vries (KdV) and modified Korteveg de Vries (mKdV) equations. The resulting equations, RKdV (reciprocal KdV) and RmKdV (reciprocal mKdV), are connected through a transformation that combines both Miura and reciprocal transformations. Lax pairs for RKdV and RmKdV are straightforwardly obtained by means of the aforementioned reciprocal transformations.
We have also identified the classical Lie symmetries for the Lax pairs of RKdV and RmKdV. Non-trivial similarity  reductions are computed and they yield non-autonomous ordinary differential equations (ODEs), whose Lax pairs are obtained as a consequence of the reductions.
\end{abstract}




\vskip 0.5cm

\section{Introduction}\label{sec:1}

Hodograph transformations are transformations involving
the interchange of dependent and independent variables \cite{clarkson,rogersshadwick}. When the
variables are swapped, the space of independent variables is called the reciprocal
space.  
Reciprocal transformations share this definition with hodograph transformations, but the former impose further requirements. Reciprocal transformations require the use of
conservative forms together with the fulfillment of specific properties
\cite{rogers4,rogers5,Est1,Est2}.

A very important advantage  when
dealing with  reciprocal transformations
is that many integrable equations 
happen to be related via reciprocal transformations. Two apparently
unrelated equations \cite{h00} or even two complete hierarchies of partial differential equations (PDEs) \cite{estevez05-1,EstSar1}  that are linked
via reciprocal transformation, are versions of a unique problem. Reciprocal transformations were proven to be a useful instrument to transform equations with peakon solutions into equations that are integrable in the Painlev\'e sense \cite{Est2,degasholm}.  For example, the composition of a Miura transformation \cite{AbloKruskalSegur,Sakovich} and a reciprocal transformation gives rise to the so called {Miura-reciprocal transformations} \cite{EstSar1} that helps us relate two different hierarchies of differential equations. 

A second significant advantage of reciprocal transformations is their
utility in the identification of integrable PDEs which cannot be identified through 
algebraic tests such as the Painlev\'e test \cite{weiss}. Our conjecture is that if an equation is integrable, there must be a
transformation that turns the initial equation into a new one for
which the Painlev\'e test is successful \cite {Est2,estevez05-1}. Similar ideas have been explored by Sakovich in \cite{sak1,sak2,sak3}.

A third bonus when using  reciprocal transformations is their role in the derivation of Lax pairs. Although it is not always possible to find a Lax pair for a given equation, a reciprocal transformation can turn it into a different one whose Lax pair is known. Therefore, by undoing the reciprocal transformation in the Lax pair of the transformed equation, we can retrieve the Lax pair of the initial equation.

This paper is an explicit display of the previous statements when applied to the famous KdV and mKdV equations \cite{AbloClark}. Nevertheless, we are dealing with the associated linear problems of these equations instead of the equations themselves. While the spectral problems for KdV and mKdV are very well known, the Lax pairs of their reciprocal versions are unknown. Furthermore, the Painlev\'e test cannot be applied to RKdV and RmKdV. Despite this, the reciprocal transformations can be extended to the eigenfunctions of the Lax pair, and the spectral problems for RKdV and RmKdV can be obtained, as described in Sections \ref{s2} and \ref{s3}. 

As KdV and mKdV are connected through a Miura transformation, we can conjecture that something similar should happen with their reciprocal versions. In section \ref{s4}, we have successfully combined reciprocal transformations with the Miura transformations between  KdV and mKdV. The result is a reciprocal-Miura transformation between  RKdV and RmKdV. Section \ref{s4} is devoted to this subject.

Finally, in Section \ref{s5}, we shall calculate  the classical Lie point symmetries for the spectral problems of RKdV and RmKdV, leading to a three parameter symmetry group in each case. We have identified some non-trivial reductions which yield the reduced ODE together with its associated spectral problem. The identification of the symmetries of the Lax pair implies the consideration of the spectral parameter as an independent variable according to ideas expressed by different authors \cite{levi2,cies1,marv3,EstGand}.

\section {Reciprocal transformations for mKdV and its spectral problem}\label{s2}
As it is well known  \cite{AbloClark}, the celebrated mKdV equation
\begin{equation}
    u_t+u_{xxx}-6u^2u_x=0\label{eq1}
\end{equation}
has a two-component Lax pair which can be written in matrix form as 

\begin{equation}
\begin{aligned}
& \vec\psi_x=b\,\vec\psi\\
& \vec\psi_t=c\,\vec\psi\label{eq2}
\end{aligned}
\end{equation}

\noindent
where $\Lambda$ is the spectral parameter and $\vec\psi=(\psi_1,\psi_2)^\intercal$ is a two-component eigenfunction. 
Furthermore, $b$ and $c$ are the $2\times2$ matrices

\begin{equation}
\begin{aligned}
 &b=\left( \begin{array}{cc} -u&\Lambda \\ \Lambda & u
		\end{array}
		\right)\\
		&c=\left( \begin{array}{cc} u_{xx}-2u^3+4\Lambda^2u&2\Lambda \left[u^2-u_x-2\Lambda^2 \right]\\2\Lambda\left[u^2+ u_x-2\Lambda^2\right] &-u_{xx}+2u^3-4\Lambda^2u
		\end{array}\label{eq3}
		\right)
		\end{aligned}
\end{equation}

Equation (\ref{eq1}) can be written in conservative form as
\begin{equation}
    u_t=\left[2u^3-u_{xx}\right]_x\label{eq4}
\end{equation}
which allows us to introduce the following exact derivative
\begin{equation}
   dz=u\,dx+\left[2u^3-u_{xx}\right]dt\label{eq5}
\end{equation}

Therefore, a reciprocal transformation can be introduced as
\begin{equation}
 x=x(z,\tau)\label{eq6}
\end{equation}
where

\begin{eqnarray}
&&dx=\frac{dz}{u}-\frac{2u^3-u_{xx}}{u}\,d\tau\nonumber \\
		&&t=\tau\label{eq7}
\end{eqnarray}
Equation (\ref{eq7})  implies that the derivatives of $x(z,\tau)$ are
\begin{eqnarray}
&&x_z=\frac{1}{u}\nonumber\\
&&	x_{\tau}=\frac{u_{xx}}{u}-2u^2\label{eq8}
\end{eqnarray}

Hence, the reciprocal transformation (\ref{eq7}), when applied to the derivatives, reads

\begin{equation}
\begin{aligned}
&\frac{\partial}{\partial x}=u\frac{\partial}{\partial  z} \\
		&\frac{\partial}{\partial t}=\frac{\partial}{\partial \tau}+\left[2u^3-u_{xx}\right]\frac{\partial}{\partial z}\label{eq9}
\end{aligned}
\end{equation}
If we define
\begin{equation}
u(x,t)=U(z,\tau)\label{eq10}
\end{equation}
we have
\begin{eqnarray}
&&u_x=UU_z \nonumber\\
		&&u_{xx}=U^2U_{zz}+UU_z^2\label{eq11}
\end{eqnarray}
and Equations (\ref{eq8}) provide
\begin{eqnarray}
&&x_z=\frac{1}{U}\nonumber\\
&&	x_{\tau}=UU_{zz}+U_z^2-2U^2 \label{eq12}
\end{eqnarray}

The cross derivatives of (\ref{eq12}) yield
$$\left[\frac{1}{U}\right]_{\tau}=\left[UU_{zz}+U_z^2-2U^2\right]_z $$
and finally
\begin{equation}U_{\tau}+\left[U^3\left(U_{zz}-U\right)\right]_z=0\label{eq13}\end{equation}

Equation (\ref{eq13}) shall be denoted RmKdV in the future because it is the reciprocal version of mKdV (\ref{eq1}) obtained through the reciprocal transformation (\ref{eq7}) and (\ref{eq10}).

\subsection{Spectral problem of RmKdV}
An interesting remark is that Equation (\ref{eq13}) should be integrable, due to its direct relation via a reciprocal transformation to an integrable one. Nevertheless, integrability tests such as those based in the Painlev\'e property \cite{weiss} do not work for these equation because a leading index cannot be even identified. Similar problems appear for some well-known integrable equations such as Camassa-Holm \cite{CH} or  Degasperis-Procesi \cite{degasholm}. In previous papers of one of the authors \cite{Est1,Est2,estevez05-1}, reciprocal transformations have been presented as an efficient method to identify the spectral problem of an equation without Painlev\'e property by means of its reciprocal relation to an equation which successfully passes the Painlev\'e test. This same procedure can be applied in order to get a Lax pair for Equation (\ref{eq13}). Actually, we can introduce a transformed eigenvector defined as
\begin{equation}
   \vec \Psi(z,\tau)=\vec\psi(x,t)
\end{equation}

By using the transformation of the derivatives given in (\ref{eq9}) we easily obtain the following Lax pair

\begin{equation}
\begin{aligned}
 & \vec\Psi_z=B\,\vec\Psi\\
 & \vec\Psi_{\tau}=\Lambda C\,\vec\Psi\label{eq15}
\end{aligned}
\end{equation}

where $B$ and $C$ are the $2\times2$ matrices

\begin{equation}
\begin{aligned}
 &B=\left( \begin{array}{cc} -\displaystyle{1}&\displaystyle{\frac{\Lambda}{U}} \\ \displaystyle{\frac{\Lambda}{U}} & 1
		\end{array}
		\right)\\
		&C=\left( \begin{array}{cc}4\Lambda U&UU_{zz}+U_z^2-2UU_z-4\Lambda^2\\UU_{zz}+U_z^2+2UU_z-4\Lambda^2 &-4\Lambda U
		\end{array}
		\right) \label{eq16}
\end{aligned}
\end{equation}

It is easy to check that the compatibility condition $\Psi_{z\tau}=\Psi_{\tau z}$ between the two \mbox{components of (\ref{eq15})} is precisely Equation (\ref{eq13}).

\section {Reciprocal transformations for KdV and its spectral problem}\label{s3}
Let 
\begin{equation}
m_t+m_{xxx}-6mm_x=0\label{eq16c}
\end{equation}
be the well-known Korteveg de Vries equation \cite{AbloClark}.
Equally renowned is the existence of the Miura transformation
\begin{equation}
m_1=u^2- u_x,\quad m_2=u^2+ u_x \label{eq17}
\end{equation}
between two solutions $m_1,\,m_2$ of (\ref{eq16c}) and a solution $u$ of (\ref{eq1}).
Besides that, each of the components of the eigenvector  $\vec\psi=(\psi_1,\psi_2)^\intercal$ of mKdV satisfies the KdV scalar  Lax pair
\begin{equation}
\begin{aligned}
&(\psi_i)_{xx}=(m_i-\lambda)\psi_i\\
&(\psi_i)_t=(2m_i+4\lambda)(\psi_i)_x-(m_i)_x\psi_i\\
\end{aligned}
\label{eq18}
\end{equation}
for $i=1,\,2$ and where
$$\lambda=-\Lambda^2$$

Equation (\ref{eq16c}) can be written in conservative form as 
\begin{equation}
m_t=\left[3m^2-m_{xx}\right]_x\label{eq19}
\end{equation}

The conservative Equation (\ref{eq19}) allows us to introduce the exact derivative
\begin{equation}
dy=mdx+\left[3m^2-m_{xx}\right]dt\label{eq20}
\end{equation}
that yields a reciprocal transformation 
\begin{equation}
x=x(y,T)\label{eq21}
\end{equation}
where
\begin{equation}
\begin{aligned}
&dx=\frac{dy}{m}-\frac{3m^2-m_{xx}}{m}\,dT \\
		&T=t
\end{aligned}\label{eq22}
\end{equation}

The induced transformation of the derivatives is
\begin{equation}
\begin{aligned}
&\frac{\partial}{\partial x}=m\frac{\partial}{\partial  y} \\
		&\frac{\partial}{\partial t}=\frac{\partial}{\partial T}+\left[3m^2-m_{xx}\right]\frac{\partial}{\partial y}\label{eq23}
\end{aligned}
\end{equation}

If we define
\begin{equation}
m(x,t)=M(y,T)\label{eq24}
\end{equation}
its derivatives are
\begin{eqnarray}
&&m_x=MM_y \nonumber\\
		&&m_{xx}=M^2M_{yy}+MM_y^2\label{eq25}
\end{eqnarray}
and, from Equation (\ref{eq22}), it is possible to write
\begin{eqnarray}
&&x_y=\frac{1}{M}\nonumber\\
		&&x_T=MM_{yy}+M_y^2-3 M\nonumber
\end{eqnarray}
whose compatibility condition $x_{yT}=x_{Ty}$ yields
\begin{equation}
\left[\frac{1}{M}\right]_T=\left[MM_{yy}+M_y^2-3M\right]_y\nonumber
\end{equation}
or, in a more compact form,
\begin{equation}
\left[M\right]_T+\left[M^3\left(M_{yy}-1\right)\right]_y=0\label{eq26}
\end{equation}

We shall denote Equation (\ref{eq26}) as RKdV due to the fact that it is the reciprocal version of the KdV equation.

\subsection{Reciprocal transformations for the spectral problem of KdV}
Let us write again the Lax pair (\ref{eq18}) for KdV
\begin{eqnarray}
&&\psi_{xx}=(m-\lambda)\psi\nonumber\\
&&\psi_t=(2m+4\lambda)\psi_x-m_x\psi\label{eq27}
\end{eqnarray}
where the subscripts have been removed for simplicity.
The eigenfunction in the new variables could be introduced as 
\begin{equation}
\psi(x,t)=\Phi(y,T)\label{eq28}
\end{equation}

By using (\ref{eq23}) and  (\ref{eq24}) in  (\ref{eq27}), we get the Lax pair for the reciprocal version (\ref{eq26}) of KdV. This Lax pair reads
\begin{eqnarray}
  &&\Phi_{yy}=-\frac{M_y}{M}\,\Phi_y+\left[\frac{1}{M}-\frac{\lambda}{ M^2}\right]\Phi\nonumber\\
&& \Phi_{T}=\left(M^2 M_{y y}+MM_y^2-M^2+4\lambda M\right)\Phi_y-MM_y\Phi
\label{eq29}
\end{eqnarray}

Notice that the Painlev\'e test cannot be applied to (\ref{eq26}). Nevertheless, the corresponding reciprocal transformation to an equation with the Painlev\'e property, such as (\ref{eq16c}), yields the appropriate Lax pair.

\section {Reciprocal-Miura transformations between  RKdV and RmKdV}\label{s4}
As we said before, Equations (\ref{eq1}) and (\ref{eq16c}) are related through the Miura \linebreak transformation \cite{AbloKruskalSegur,Sakovich}, 

\begin{equation}
m=u^2+ku_x,\quad k=\pm 1\label{eq30}
\end{equation}

The reciprocal transformations introduced in the previous sections imply that
\begin{equation}
u(x,t)=U(z,\tau),\quad m(x,t)=M(y,T)\label{eq31}
\end{equation}
where the transformations among the independent variables are
\begin{subequations}
\begin{eqnarray}
  &&  dx=\frac{dz}{u}-\frac{2u^3-u_{xx}}{u}\,d\tau \label{eq32a}\\
&&dx=\frac{dy}{m}-\frac{3m^2-m_{xx}}{m}\,dT \label{eq32b}\\
&&   t=\tau=T
\end{eqnarray}\label{eq32}
\end{subequations}

A question immediately arises, is there any sort of Miura relation between $M(y,T)$ and $U(z,\tau)$?. Let us proceed to answer this question. By using (\ref{eq11}) and (\ref{eq31}) in (\ref{eq30}), we have
\begin{equation}
M=U\left(U+kU_z\right)\label{eq33}
\end{equation}

Equation (\ref{eq33}) can be considered to be the reciprocal version of the Miura map (\ref{eq30}). Nevertheless, it should be combined with the transformation between the set of variables $\left\{z,\tau\right\}$ and $\left\{y,T\right\}$, defined in \eqref{eq32a} and \eqref{eq32b}. The combination of these transformations gives rise to
\begin{eqnarray}
  && dy= \frac{m}{u}dz+\left[3m^2-m_{xx}+\frac{m}{u}\left(u_{xx}-2u^3\right)\right]\,d\tau 
 \nonumber\\
&& T=  \tau \label{eq34}
\end{eqnarray}

This result allows us to write a \textbf{Miura-reciprocal transformation} between  \mbox{RmKdV (\ref{eq13})} and RKdV (\ref{eq26}) as
\begin{equation}
\begin{aligned}
  &M(y,T)=U(z,\tau)\left(U(z,\tau)+kU_z(z,\tau)\right)\\
  &dy=\left[U+kU_{z}\right]dz+\left[(U^4-U^3U_{zz})+k(U^4-U^3U_{zz})_z\right]d\tau\\
  &T=\tau
\end{aligned}\label{eq36}
\end{equation}

\begin{figure}[H]
\centering
\hspace{3cm}
$
\xymatrix{*+<1cm>[F-,]{\begin{matrix}\text{{   KdV  }}\end{matrix}
} \ar[rrrrr]^{\text{Reciprocal transf. }\eqref{eq22}\text{ and }\eqref{eq24}} \ar@2{<->}[d]^{\text{Miura transf. }\eqref{eq30}} &  & & & &*+<1cm>[F-,]{\begin{matrix}\text{{  RKdV  }}\\\end{matrix}}\ar@2{<->}[d]^{\text{\textbf{Miura-reciprocal transf. }\eqref{eq36}}}\\  *+<1cm>[F-,]{\begin{matrix}\text{{mKdV}}\end{matrix}} \ar[rrrrr]^{\text{Reciprocal transf. }\eqref{eq7}\text{ and }\eqref{eq10}} &  & & & &*+<1cm>[F-,]{\begin{matrix}\text{{RmKdV}}\\\end{matrix}}}
$
\caption{Miura-reciprocal transformation for RKdV and RmKdV.}
\label{Fig1}
\end{figure}
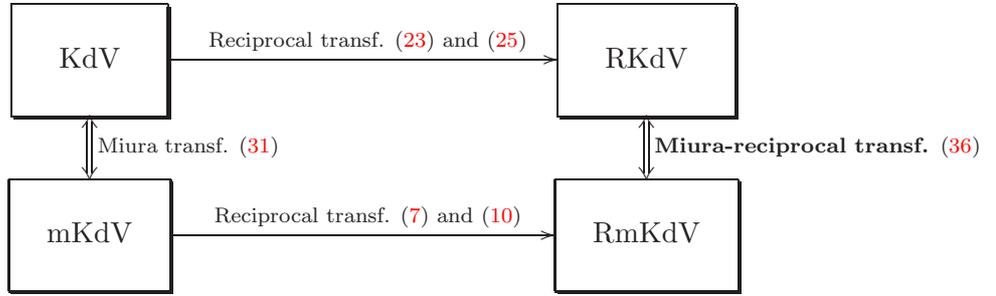

The whole picture concerning the relations among the integrable equations and their associated spectral problems analyzed in this article can be illustrated in Figure \ref{Fig1}.
The results obtained in the previous sections may be summarized in the following theorem and corollary:

\begin{theorem}\label{thm1}
Let u(x,t) be a solution of the mKdV equation (\ref{eq1}) and m(x,t) be a solution of the KdV equation (\ref{eq16c}), related by the Miura map (\ref{eq30}). The conservative form of KdV and mKdV allows the introduction of the reciprocal transformations 
\begin{align}
dx&=\frac{dz}{u}-\frac{2u^3-u_{xx}}{u}\,d\tau,
&t&=\tau,&u(x,t)&=U(z,\tau)\label{r1}\\
dx&=\frac{dy}{m}-\frac{3m^2-m_{xx}}{m}\,dT,
&t&=T,&m(x,t)&=M(y,T)\label{r2}
\end{align}
such that (\ref{r1}), when applied to mKdV, provides the reciprocal mKdV (RmKdV) equation
\begin{equation}
U_{\tau}+\left[U^3\left(U_{zz}-U\right)\right]_z=0
\label{eqth1}
\end{equation}
whilst (\ref{r2}) gives rise to the reciprocal version of KdV (RKdV),
\begin{equation}
\left[M\right]_T+\left[M^3\left(M_{yy}-1\right)\right]_y=0
\label{eqth2}
\end{equation}

Then, the Miura-reciprocal transformation defined by
\begin{equation}
\begin{aligned}
  &M(y,T)=U(z,\tau)\left(U(z,\tau)+kU_z(z,\tau)\right)\\
  &dy=\left[U+kU_{z}\right]dz+\left[(U^4-U^3U_{zz})+k(U^4-U^3U_{zz})_z\right]d\tau\\
  &T=\tau
\end{aligned},\qquad k=\pm 1
\label{eqth3}
\end{equation}
establishes a map between RmKdV (\ref{eqth1}) and RKdV (\ref{eqth2}).
\end{theorem}
\begin{corollary}\label{cor1}
Reciprocal transformations (\ref{r1}) and (\ref{r2}) induce a transformation over the spectral problems of mKdV (\ref{eq2}) and KdV (\ref{eq27}), respectively. Then, the resulting Lax pairs for RmKdV (\ref{eq15}) and RKdV (\ref{eq29}) are linked via the map (\ref{eqth3}) such that each component of the eigenvector $\vec{\Psi}(z,\tau)=(\Psi_1,\Psi_2)^\intercal$ in (\ref{eq15}) satisfies the scalar linear problem for $\Phi(y,T)$ in (\ref{eq29}). 
\end{corollary}

\section{Lie symmetries for the transformed spectral problems}\label{s5}
As the final section of this paper, we identify the Lie point symmetries \cite{BK,Olver,Lie} for the spectral problems given in (\ref{eq15}) and (\ref{eq29}).
Lie symmetries were first introduced by Lie \cite{Lie} in order to solve ordinary differential equations (ODEs) or reduce a system
of equations to a simpler form \cite{Steph,wint}. Lie point symmetries represent a powerful tool, although they involve lengthy
calculations for the most part.  Each Lie point
symmetry leads us to a reduced version of the equation with the number of independent variables diminished by one. This
fact implies that any solution of a PDE can be derived by its simmilarity  reduction  to an ODE. 
Much less frequent is the identification of Lie symmetries of the spectral problem of a PDE system \cite{leg}. The
introduction of an spectral (or non-isospectral) parameter in a linear problem without spectral parameter through one-parameter
groups of Lie point symmetries was proposed by Levi {\it et al.} \cite{levi2,levi1}. This group interpretation of the spectral
parameter has been extensively studied by Cieśliński  {\it et al.} in several papers \cite{cies1,cies2,cies3} and by Marvan  {\it et al.} \cite{marv3,marv1,marv2}. Lie symmetries for the Lax pair \cite{EstGand} of a PDE provide much more information than just the symmetries of the PDE itself since they allow to get the similarity reductions of both the PDE and the spectral problem \cite{BGEP}. This approach requires to consider the spectral parameter as an independent variable and the eigenfunctions as fields \cite{AlEstCon,AlEstLej}.
\subsection{Lie point symmetries for RKdV}
Let us consider a one-parameter Lie  group \cite{Steph} of infinitesimal transformations of the independent variables $\left\{y,T\right\}$, one dependent field  $M(y,T)$, the spectral parameter $\lambda$ and the eigenfunction $\Psi(y, T, \lambda)$, given by
\begin{equation}
    \begin{aligned}
    \tilde{y}&=y+\epsilon\,\xi_y(y,t,M,\lambda,\Phi)\\
    \tilde{T}&=T+\epsilon\,\xi_T(y,t,M,\lambda,\Phi)\\
    \tilde{\lambda}&=\lambda+\epsilon\,\xi_\lambda(y,t,M,\lambda,\Phi)\\
    \tilde{M}&=M+\epsilon\,\eta_{M}(y,t,M,\lambda,\Phi)\\
    \tilde{\Phi}&=\Psi+\epsilon\,\eta_{\Phi}(y,t,M,\lambda,\Phi)
    \label{eq42}
    \end{aligned}
\end{equation}
where $\epsilon$ is the group parameter. The associated vector field that generates these infinitesimal transformations reads 
\begin{equation}\label{eq37}
X_{\text{RKdV}} = \xi_y \frac{\partial}{\partial y} + 
\xi_T\frac{\partial}{\partial T}+
\xi_\lambda \frac{\partial}{\partial \lambda}+
\eta_M \frac{\partial}{\partial M}+
\eta_{\Phi} \frac{\partial}{\partial \Phi}
\end{equation}

This infinitesimal transformation induces a well known one in the derivatives of the fields and it  must preserve the invariance of the starting system of PDEs. By applying Lie's method \cite{BK,Olver,Lie}, this procedure yields an overdetermined system of PDEs for the infinitesimals called the determining equations, whose solutions provide the classical symmetries.  We have  computed these symmetries with the help of Maple for the Lax pair given in (\ref{eq29}). The result is
\begin{equation}
    \begin{aligned}
    \xi_y&=C_1y+C_2\\
    \xi_T&=-3C_1T+C_3\\
    \eta_M&=2C_1\,M\\
    \xi_{\lambda}&=2C_1\,\lambda\\
    \eta_{\Phi}&=f(\lambda)\Phi
    \end{aligned}\label{eq38}
\end{equation}
where $C_i, i=1...3$ are arbitrary constants. $f(\lambda)$ is an arbitrary function of $\lambda$ which is a consequence  of the invariance of the Lax pair (\ref{eq29}) when $\Phi$ is multiplied by an arbitrary function of $\lambda$.  Symmetries (\ref{eq38}) yield the three-dimensional Lie algebra spanned by $\langle y\partial_y-3T\partial_T+2\lambda\partial_{\lambda}+2M\partial_M,\partial_y,\partial_T\rangle$ and the infinite-dimensional subalgebra $\langle f(\lambda)\partial_{\Phi}\rangle$ due to the presence of the arbitrary function $f(\lambda)$ \cite{Olver,wint}. For each parameter we can have a similarity reduction. We shall identify the reduction associated to $C_1$ because the symmetries  connected to $C_2$ and $C_3$ are no more than translations which yield \mbox{trivial reductions.}

\subsubsection*{Reduction associated to $C_1$}

The vector field associated to $C_1$ is 
\begin{equation}
X_1 = y \frac{\partial}{\partial y} 
-3T\frac{\partial}{\partial T}+
2\lambda\frac{\partial}{\partial \lambda}+
2M\frac{\partial}{\partial M}
\end{equation}
and the reduction can be obtained through the integration of the characteristic system
\begin{equation} \label{characteristic}
\frac{dy}{y}=\frac{dT}{-3T}=\frac{d\lambda}{2\lambda}=\frac{d\,M}{2M}=\frac{d\Phi}{0}
\end{equation}

This integration yields
\begin{itemize}
\item Reduced independent variable
$$ Y=yT^{1/3}$$
\item Reduced field
$$N(Y)=M(y,T)T^{2/3}$$
\item Reduced spectral parameter
$$ \beta =\lambda T^{2/3}$$
where $\beta$ is the new spectral parameter.
\item Reduced eigenfunction
$$\overline{\Phi}(Y,\beta)=\Phi(y,T)$$
\item Reduced Lax pair
\begin{eqnarray}
  &&\overline{\Phi}_{YY}=-\frac{N_Y}{N}\,\overline{\Phi}_{Y}+\frac{N-\beta}{N^2}\,\overline{\Phi}\nonumber\\&&\beta\overline{\Phi}_{\beta}=\left[\frac{3}{2}\left(N^2N_{YY}+NN_Y^2-N^2\right)+6\beta N-\frac{Y}{2}\right]\overline{\Phi}_{Y}-\frac{3}{2}NN_Y\overline{\Phi}_{Y}\nonumber
\end{eqnarray}

\item Reduced ODE
$$\left[N^3(N_{YY}-1)\right]_Y+\frac{Y}{3}N_Y-\frac{2N}{3}=0$$

\end{itemize}

We should remark that this is an elegant way of obtaining a rather non-trivial Lax pair for a non-autonomous ODE. The introduction of the spectral parameter $\beta$ arises in a natural form.
\subsection{Lie point symmetries for RmKdV}
Let us consider now the spectral problem obtained in (\ref{eq15}). In this case, we should consider  Lie  infinitesimal transformations for the independent variables $\left\{z,\tau\right\}$, one dependent field  $U(z,\tau)$, the spectral parameter $\Lambda$ and the eigenfunctions $\vec\Psi(z,\tau, \Lambda)=(\Psi_1,\,\Psi_2)^\intercal$, given by
\begin{equation}
    \begin{aligned}
    \tilde{z}&=z+\epsilon\,\xi_z(z,\tau,U,\Lambda,\Psi_1,\Psi_2)\\
    \tilde{\tau}&=\tau+\epsilon\,\xi_{\tau}(z,\tau,U,\Lambda,\Psi_1,\Psi_2)\\
    \tilde{\Lambda}&=\Lambda+\epsilon\,\,\xi_{\Lambda}(z,\tau,U,\Lambda,\Psi_1,\Psi_2)\\
    \tilde{U}&=U+\epsilon\,\eta_{U}(z,\tau,U,\Lambda,\Psi_1,\Psi_2)\\
    \tilde{\Psi_1}&=\Psi_1+\epsilon\,\eta_{\Psi_1}(z,\tau,U,\Lambda,\Psi_1,\Psi_2)\\
        \tilde{\Psi_2}&=\Psi_2+\epsilon\,\eta_{\Psi_2}(z,\tau,U,\Lambda,\Psi_1,\Psi_2)
    \end{aligned}
\end{equation}

Lie symmetries have been computed  using Maple, resulting in

\begin{equation}
    \begin{aligned}
    \xi_z&=D_2\\
    \xi_{\tau}&=3D_1{\tau}+D_3\\
    \eta_U&=-D_1\,U\\
    \xi_{\Lambda}&=-D_1\,\Lambda\\
    \eta_{\Psi_1}&=g(\Lambda)\Psi_1\\
    \eta_{\Psi_2}&=g(\Lambda)\Psi_2
    \end{aligned}
\end{equation}

The most interesting reduction arises from the one associated to $D_1$, since parameters $D_2,\,D_3$ provide trivial reductions.
\subsubsection*{Reductions associated to $D_1$}
The characteristic system  for the symmetry related to $D_1$ is
\begin{equation} 
\frac{dz}{0}=\frac{d\tau}{3\tau}=\frac{d\Lambda}{-\Lambda}=\frac{d\,U}{-U}=\frac{d\Psi_1}{0}=\frac{d\Psi_2}{0}
\end{equation}
whose integration yields the following reduction
\begin{itemize}
\item Reduced independent variable
$$ Z=z$$
\item Reduced field
$$H(Z)=U(z,\tau)\tau^{1/3}$$
\item Reduced spectral parameter
$$ \alpha=\Lambda \tau^{1/3}$$
where $\alpha$ is the new spectral parameter for the reduced spectral problem.
\item Reduced eigenfunctions
$$\Omega_1(Z,\alpha)=\Psi_1(z,\tau)$$
$$\Omega_2(Z,\alpha)=\Psi_2(z,\tau)$$
denoted in the following as $\vec{\Omega}=(\Omega_1,\Omega_2)^\intercal$.
\item Reduced Lax pair
\begin{eqnarray}
 && \vec{\Omega}_Z=\hat B\vec{\Omega}\nonumber\\
 && \vec{\Omega}_{\alpha}=3\hat C\vec{\Omega}\nonumber
\end{eqnarray}
where $\hat B$ and $\hat C$ are the $2\times 2$ matrices

\begin{eqnarray}
 &&\hat B=\left( \begin{array}{cc} -1&\displaystyle{\frac{\alpha}{H}} \\ \displaystyle{\frac{\alpha}{H}} & 1
		\end{array}
		\right)\nonumber \\
		&&\hat C=\left( \begin{array}{cc}4\alpha H &HH_{ZZ}+H_Z^2-2HH_Z-4\alpha^2\\HH_{ZZ}+H_Z^2+2HH_Z-4\alpha^2 &-4\alpha H
		\end{array}
		\right)\nonumber
\end{eqnarray}

\item Reduced ODE
$$\left[H^3(H_{ZZ}-H)\right]_Z-\frac{H}{3}=0$$

\end{itemize}

As in the case of RKdV, the spectral problem of the reduced ODE is nicely obtained through the introduction of a spectral parameter $\alpha$ which arises from the reduction of the spectral parameter $\Lambda$ of the former Lax pair.
\section{Conclusions}\label{s6}
In this paper we have presented reciprocal transformations for the spectral problems of KdV and mKdV. The link between the resulting reciprocal equations is established by means of the so-called Miura-reciprocal transformation, which stands for a combination of the proposed reciprocal transformations and the Miura map inherited from the existing relation between KdV and mKdV. These results are compiled in Theorem \ref{thm1} and \mbox{Corollary \ref{cor1}.}

Furthermore, the classical Lie symmetries for the reciprocal version of these spectral problems are identified. These symmetries may straightforwardly yield non-trivial similarity reductions. When dealing with the Lax pair instead of just the equation itself, we also get the reduced spectral problem for the reduced ODE. The reduction of the spectral parameter plays an important role in this process.

\vspace{0.2cm}

\textbf{Author contributions}: The authors have contributed equally to this work. All authors have read and agreed to the published version of the manuscript. 

\textbf{Funding}: This research has been supported in part by MICINN (Grant PID2019-106820RBC22) and Junta de Castilla y Le\'on (Grant SA256P18). P. Albares also acknowledges support from the predoctoral grant FPU17/03246.

\textbf{Institutional Review Board Statement}: Not applicable.

\textbf{Informed Consent Statement}: Not applicable.

\textbf{Data Availability Statement}: Not applicable.

\textbf{Conflicts of Interest}: The authors declare no conflict of interest.

\end{document}